\documentclass[twocolumn,prb,showpacs,superscriptaddress]{revtex4}

\usepackage{amsmath,graphicx,natbib}

\newcommand{\V}[1]{ { \bf #1 } }

\begin{document}

\title{Phase diagram of the one-dimensional, two-channel Kondo lattice model}

\author{T.\ Schauerte}
\author{D.\;L.\ Cox}
\affiliation{Department of Physics, University of California, Davis,
  CA 95616, USA}

\author{R.\;M.\ Noack}
\affiliation{Fachbereich Physik, Philipps-Universit\"at Marburg,
  D-35032 Marburg, Germany}

\author{P.\;G.\;J.\ van Dongen}
\affiliation{Institut f\"ur Physik, Universit\"{a}t Mainz, 55099
  Mainz, Germany}

\author{C.\ D.\ Batista}
\affiliation{Theoretical Division, Los Alamos National Laboratory, Los
  Alamos, NM 87545}

\date{\small \today}

\begin{abstract}
    Employing the density matrix renormalization group method and
    strong-coupling perturbation theory, we study the phase diagram of
    the SU(2)$\times$SU(2) Kondo lattice model in one dimension.
    We show that, at quarter filling, the system can exist in two phases
    depending on the coupling strength.
    The weak-coupling phase is dominated by RKKY exchange correlations
    while the strong-coupling phase is characterized by
    strong antiferromagnetic correlations of the channel degree of freedom.
    These two phases are separated by a quantum critical point.
    For conduction-band fillings of less than one quarter, we find a paramagnetic
    metallic phase at weak coupling and a ferromagnetic phase at
    moderate to strong coupling.
\end{abstract}


\maketitle

Although Landau's theory of Fermi liquids is one of the
cornerstones of modern condensed matter physics,\cite{Landau56}
many materials show metallic properties that do not fit into this
framework. For instance, it is well known that one-dimensional
systems usually behave as Luttinger liquids rather than Fermi
liquids.\cite{TLL} In higher dimensions, the proximity of a
quantum critical point (QCP) is considered to be responsible for
the non-Fermi liquid properties of systems like
CeCu$_{6-x}$Au$_x$.\cite{Loehn96} The quadrupolar Kondo effect has
been proposed as an alternative source of non-Fermi liquid
behavior. This effect is described by the two-channel Kondo model
\cite{Nozieres80,Cox87}. This model has a non-Fermi liquid
ground-state because of frustration in the screening of a
localized impurity by two degenerate conduction electron channels
if the degeneracy of the channels $M$ is greater than twice the
impurity spin $S$.\cite{TCKM} The single-impurity two-channel
Kondo model can provide an adequate description of dilute systems
like Th$_{1-x}$U$_x$Ru$_2$Si$_2$ \cite{Amitsuka93} or
Y$_{1-x}$U$_x$Pd$_3$,\cite{Andraka91} but it does not incorporate
the lattice effects that become relevant in fully concentrated
compounds like UBe$_{13}$ \cite{Cox87}. These materials are
described by the two-channel Kondo lattice model (KLM).

In this paper, we address the question of what happens when two of
these fundamental sources of non-Fermi liquid behavior coincide.
For this purpose we study the SU(2)$\times$SU(2) Kondo lattice
model in one spatial dimension. The Hamiltonian reads
\begin{eqnarray}
H &=&   - \, t \sum_{i m \sigma} \left( c^\dag_{i m\sigma}
    c^{\phantom{\dag}}_{i+1,m\sigma} + \mbox{H.c.} \right)
    \nonumber \\
  && \quad  + \frac{1}{2}J \sum_{i m \alpha \beta} {\V{S}}_i \cdot \left(
    c^\dag_{i m \alpha}
    \text{\boldmath{$\sigma$}}^{\phantom{\dag}}_{\alpha \beta}
    c^{\phantom{\dag}}_{i m \beta} \right) \; ,
\nonumber
\end{eqnarray}
where $t>0$ is the conduction electron hopping amplitude, taken to
be the same in both bands, and $c^\dag_{i m \sigma}$
($c^{\phantom\dag}_{i m \sigma}$) creates (annihilates) an
electron on lattice site $i$, $1 \leq i \leq L$ ($L$ being the
number of lattice sites), with channel flavor $m=+$ or $-$ and
spin projection $\sigma=\uparrow$ or $\downarrow$. The Heisenberg
spin operator for the localized $f\/$-electrons
($S_i=\frac{1}{2}$) is defined by ${\V{S}}_i=\frac{1}{2}
\sum_{\alpha \beta} f^\dag_{i \alpha}
\text{\boldmath{$\sigma$}}^{\phantom{\dag}}_{\alpha \beta}
f^{\phantom{\dag}}_{i \beta}$, where the $f$-operators satisfy the
constraint $f^\dag_{i \uparrow}f^{\phantom{\dag}}_{i \uparrow} +
f^\dag_{i \downarrow}f^{\phantom{\dag}}_{i \downarrow}= 1$ and
{\boldmath{$\sigma$}} is a vector of Pauli spin matrices. The
conduction band filling is defined as $n_c=n_{c+}+n_{c-}=
(N_{c+}+N_{c-})/L$ with $N_{c\pm}$ the number of conduction
electrons in channel $m=\pm$, respectively; $n_c=1$ corresponds to
the quarter filled system. The KLM may be derived from the more
fundamental periodic Anderson model in the limit of strong Coulomb
repulsion where the Kondo coupling is usually antiferromagnetic
(AF), $J > 0$,\cite{Schrieffer66} or alternatively in the
``extended Kondo limit''.\cite{Sinjukow02} In the following, we
measure all energies in units of $t$.

Not much is known about the ground-state phase diagram of the
two-channel KLM. Tsvelik and Ventura\cite{Tsvelik00} investigated
this model using a mean-field analysis and found that at half
filling the system exists in two phases. One is dominated by RKKY
exchange interaction effects, and the other by Kondo screening. A
QCP separates these two regimes. A generalized one-dimensional
two-channel KLM with an additional Heisenberg interaction, $J_H$,
between the $f$-spins was studied by Andrei and
Orignac.\cite{Andrei00} In the limit of strong $J_H$, they find
that the system is in a superconducting phase with odd-frequency
singlet pairing of the electrons. In infinite spatial dimensions,
the ground state is characterized by superconducting or magnetic
phases, which may coexist or compete.\cite{Jarrell96}

The focus on the {\em two\/}-channel system below or at quarter
filling is also motivated by the relation to the {\em single\/}-channel
system at half filling or less.
In particular, the quarter-filled case
for the two-channel model is analogous to the half-filled case for the
single-channel model in that there is one conduction electron per
impurity spin, which leads to complete screening at strong
coupling.
In the following, we will show that the situation in the
two-channel model {\em below\/} quarter filling is qualitatively
similar to the single-channel case below half filling, while exactly
at quarter filling it is quite different.
The phase diagram of the
single-channel model
is well understood, at least
qualitatively.\cite{Troyer93,Tsunetsugu94}
In the low carrier limit,
this system displays ferromagnetic order with {\em complete}
polarization, $S_{\text{tot}}=(L-N_c)/2$, where $n_c \ll 1$.\cite{Sigrist91,Sigrist92a}
In the strong-coupling limit, the ground state is ferromagnetic for all
$n_c$.\cite{Sigrist92}
Exactly at half filling, the single-channel
model is known to be a Kondo
insulator.\cite{Tsunetsugu92,Tsunetsugu94}
This is a quantum disordered phase in which the conduction electrons are bound into
local singlet states with the impurity spins, and both the spin and
charge correlation functions decay exponentially in space and time.

\begin{figure}[htb]
    \centering  \includegraphics[clip=true,width=7.5cm]{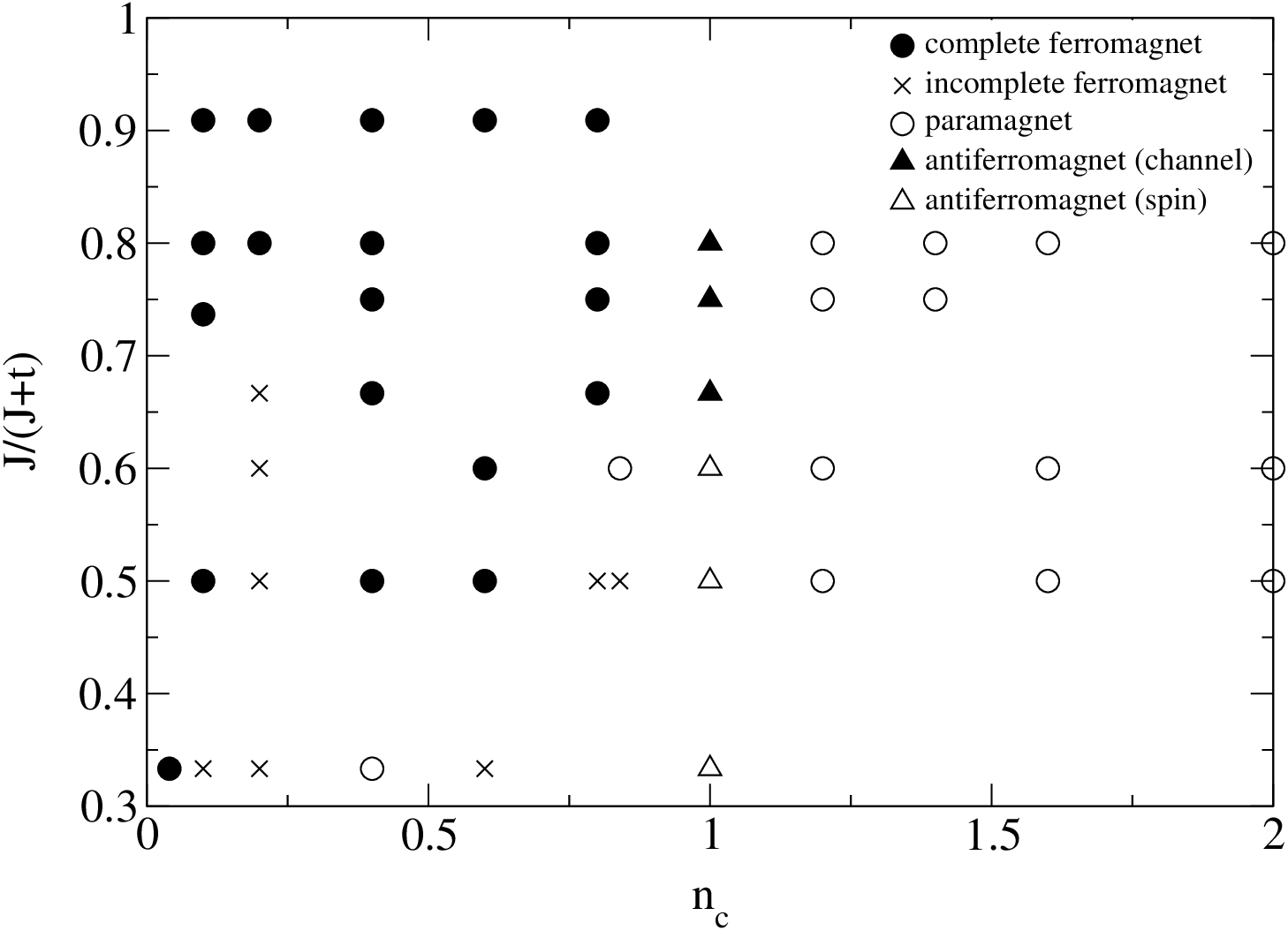}
    \caption{ Ground-state phase diagram of the two-channel KLM as
    a function of conduction band filling $n_c$.
    The AF channel- and spin-ordered phases at quarter
    filling are associated with a singlet ground state. The crosses
    indicate polarizations that extrapolate in the thermodynamic limit
    to values between 25\% and 90\% of complete polarization.
    }
\label{fig1}
\end{figure}

In order to calculate the ground-state properties of
the one-dimensional two-channel KLM, we use the finite-system
algorithm\cite{White92} of the DMRG to calculate gaps, equal-time
correlation functions, and the total spin of the ground state.
We keep up to 1000 states per block on lattices of up to $L=50$ sites
and obtain a maximum discarded weight of $10^{-5}$.
Fig.\ \ref{fig1} shows the total spin
$S_{\text{tot}}$ per site extrapolated to the thermodynamic limit
for various couplings and conduction
electron densities. Here $S_{\text{tot}}$ is
calculated directly by taking the expectation value of
${\V{S}}_{\text{tot}}^2$ in
the ground state and also by examining the degeneracy of excited
states in various $S_z$ sectors.
A finite-size extrapolation
is then used to determine whether $S_{\text{tot}} =
(L-N_c)/2$ (complete ferromagnetism), $S_{\text{tot}}
< (L-N_c)/2$ but finite (incomplete ferromagnetism), or
$S_{\text{tot}} = 0$ (paramagnetism) in the
thermodynamic limit. At quarter filling ($n_c=1$), the nature of the
ordering is also indicated for $S_{\text{tot}} = 0$ phases.

Fig.\ \ref{fig1} shows a large region of complete ferromagnetism
for $n_c < 1$ above a certain critical value $J_c$ which decreases
with decreasing $n_c$ and tends to zero as $n_c\to 0$.  In the
completely polarized phase, each conduction electron forms an
itinerant singlet with the $f$-spins. Delocalization of these
singlets leads to ferromagnetic ordering of the remaining
unscreened $f$-spins.  The complete polarization for all $J$ at
low conduction electron density is in agreement with an exact
argument \cite{Sigrist91} for a single conduction electron. For
$J<J_c$, there is a region of incomplete ferromagnetism, similar
to one that has been found in the periodic Anderson
model.\cite{Guerrero96} While we cannot rule out that this region
is due to a continuous transition to the complete ferromagnetic
phase, the local spin profiles in the incomplete ferromagnetic
regime show small ferromagnetic domains (corresponding to
polarizations between 25\% and 90\% of the complete value),
suggesting that phase separation may occur here. For $n_c \geq 1$,
we find a singlet ground state for all couplings $J$.

Exactly at quarter filling, we find two phases as a function of
$J$. At weak coupling ($J \lesssim 2.0$), the electrons of
different flavors generate independent RKKY interactions between
the localized moments. We observe strong correlations of the
$f$-spins at a wave vector $q=\pi/2$, as in the single-channel
model at {\em quarter\/} filling.\cite{Shibata96} At stronger
coupling ($J \gtrsim 2.0$), the system is in a channel AF phase,
where the correlations of the channel degree of freedom decay as
$1/r$. The two phases are separated by a QCP.

\begin{figure}[htb]
    \centering \includegraphics[clip=true,width=7.5cm]{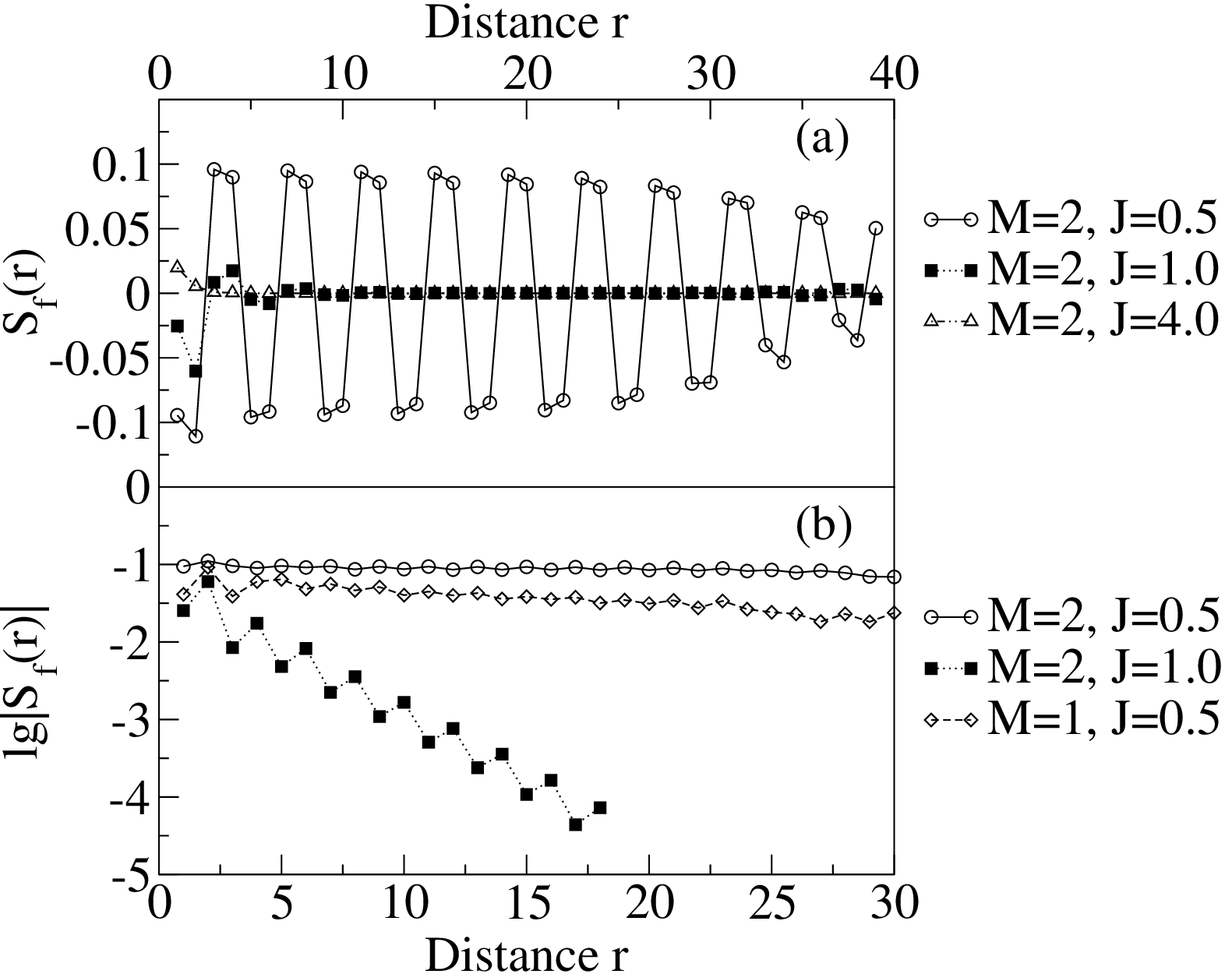}
    \caption{
    Magnetic correlation function $S_f(r)=\langle S_z(0) S_z(r)\rangle$ of
    the $f$-spins for the two-channel KLM at $n_c =1$ in (a) and
    for the single- and two-channel models, both at quarter-filling, in (b).
    }
\label{fig2}
\end{figure}

\begin{figure}[htb]
    \centering \includegraphics[clip=true,width=5.8cm]{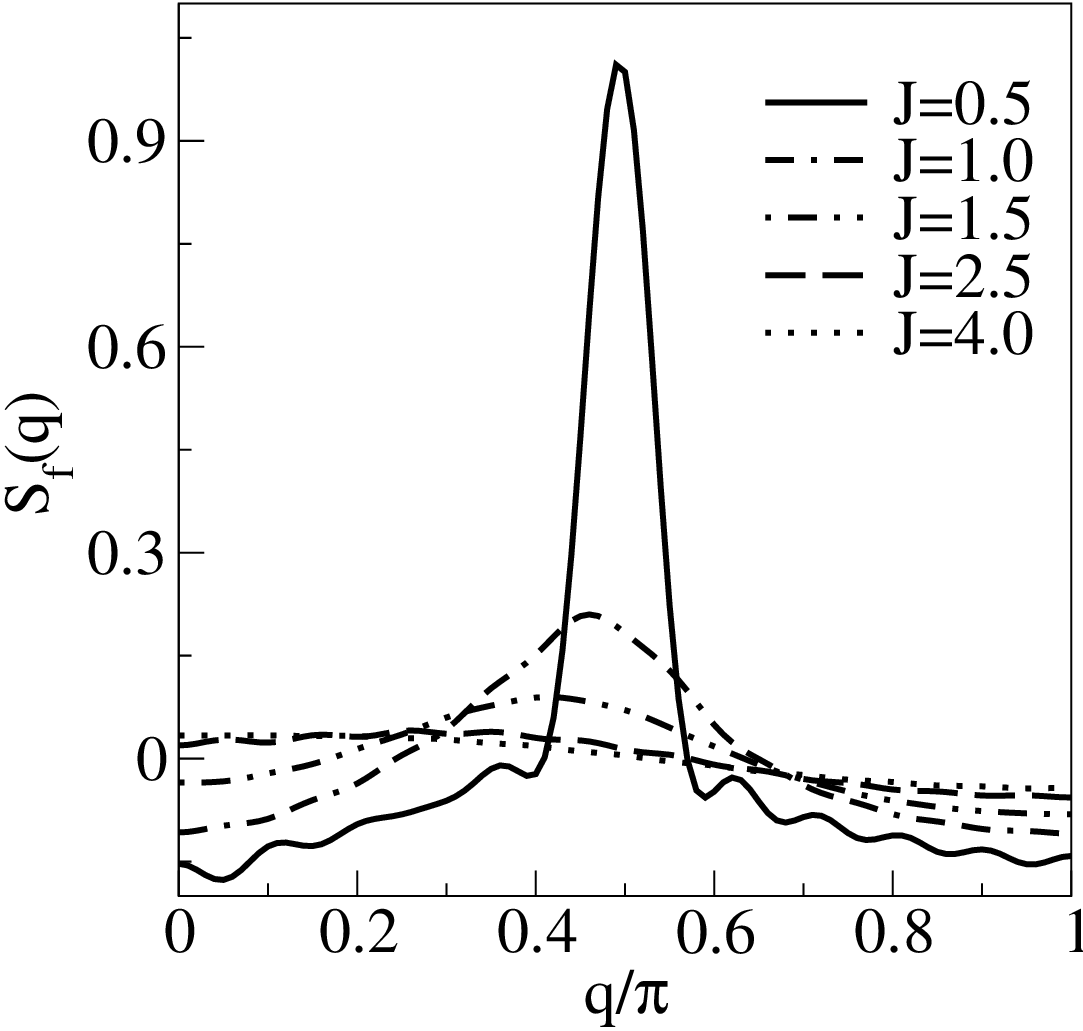}
    \caption{
    Magnetic structure factor of the two-channel KLM
    model as a function of wavevector $q$ for various $J$ at
    $n_c=1.0$.
    }
\label{fig3}
\end{figure}

Fig.\ \ref{fig2}(a) shows that in the two-channel model the
spin-spin correlation function decays slowly for weak coupling
($J=0.5\ \text{or}\ 1.0$) and is short-ranged for strong coupling
($J=4.0$), i.e., numerically zero for more than two lattice
spacings. The transition between the two behaviors occurs at $J
\approx 2.0$ (not shown in Fig.\ \ref{fig2}), where the
correlations extend over roughly two lattice sites. Fig.\
\ref{fig2}(b) shows a comparison between the spin-spin
correlations functions $S_f(r)$ of the single-channel and
two-channel KLM on a logarithmic plot. The ground state for $M=1$
in the weak-coupling limit shows RKKY liquid behavior with spatial
oscillation characterized by a wave vector $q=\pi/2$. In the $M=2$
case the correlations at $J=0.5$ decay so slowly that the
asymptotic behavior cannot be determined for the system sizes
considered, as can be seen in Fig.\ \ref{fig2}(b). At larger
$J$-values [e.g., $J=1.0$ in Fig.\ \ref{fig2}(b)], the correlation
functions appear to decay 
exponentially, consistent with the opening of a gap in the spin
excitation spectrum due to the transition into the channel-ordered
phase. Fig.\ \ref{fig3} shows the magnetic structure factor
$S_f(q)$ for the two-channel system. The amplitude of the peak at
$q=\pi/2$ (the wavevector expected for a RKKY liquid) becomes smaller with
increasing $J$, indicating that the magnetic correlations between
the $f$-spins become weaker and finally vanish at $J \approx
2.0$.\cite{FT-remark}

\begin{figure}[htb]
    \centering  \includegraphics[clip=true,width=6.2cm]{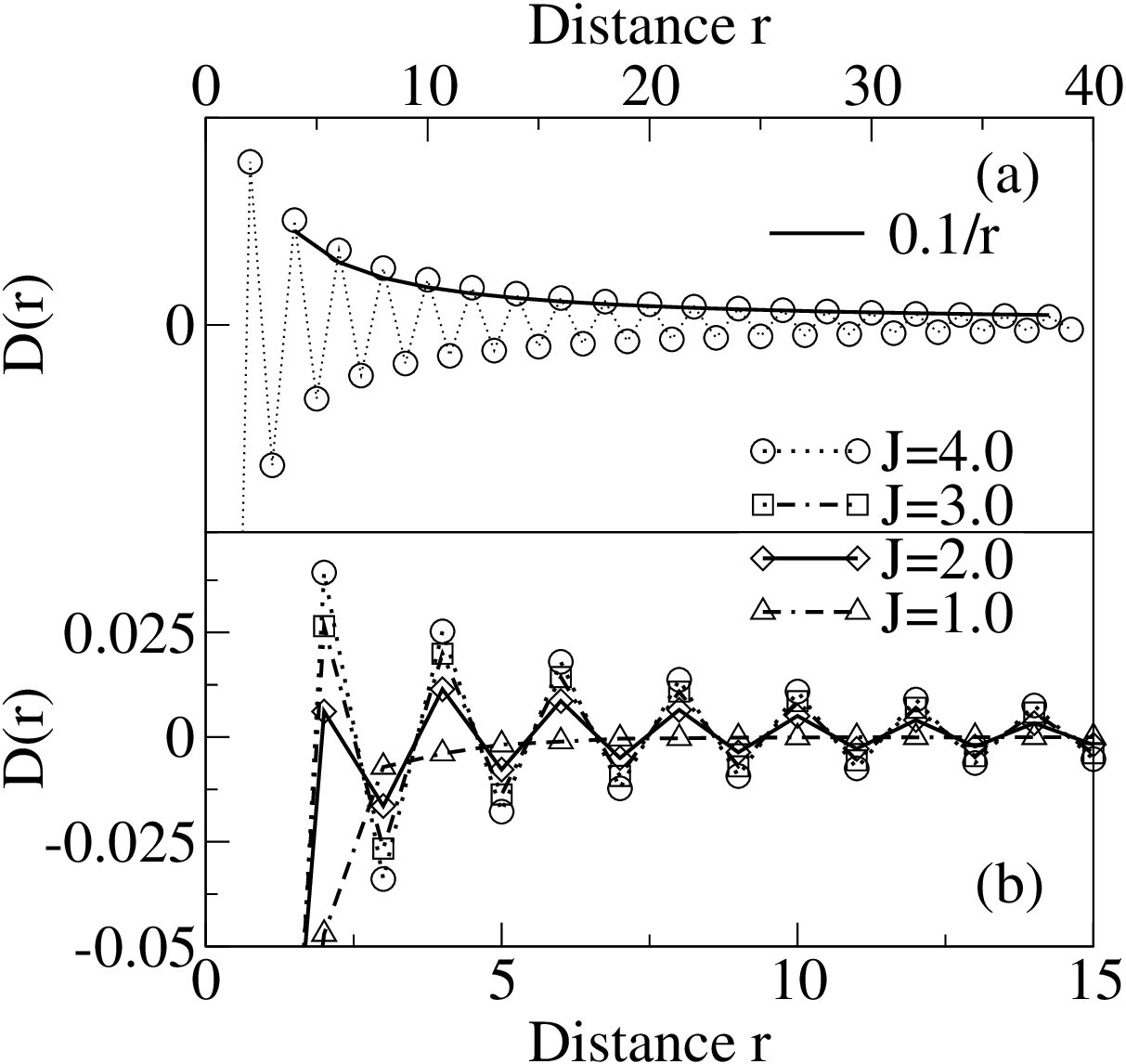}
    \caption{
      Staggered magnetization $D(r)$ for the channel degree of
      freedom for various $J$ at quarter filling ($n_c=1.0$).
      The longer distance behavior for $J=4.0$ (where there is a
      slower falloff, approximately $\propto 1/r$) is shown in (a) and
      a shorter range in $r$ is shown for all four $J$-values in (b).
    }
\label{fig4}
\end{figure}

It is known that the RKKY correlations of the single-channel KLM
at {\em quarter\/} filling become unstable with increasing
coupling toward a ferromagnetic ground
state.\cite{Tsunetsugu94,Honner97} However, the two-channel model
cannot be ferromagnetic at quarter filling
because each $f$-spin is screened by an electron,
so that some other
kind of symmetry breaking must lift the degeneracy of the ground
state. Fig.\ \ref{fig4} shows the staggered magnetization $D(r)$
of the channel degree of freedom, with
\begin{equation}
D(i-j) = \sum_{\sigma \sigma' m m'} \; m \, m' \;
       \langle \, n_{i m \sigma} \;
       n_{j m' \sigma'} \,
       \rangle \; .
\nonumber
\end{equation}
where $n_{i m\sigma}=c^\dag_{i m\sigma}c^{\phantom{\dag}}_{i m
\sigma}$. In Fig.\ \ref{fig4}(a) one sees that $D(r)$ decays with
$1/r$ as function of the distance $r=i-j$ for $J=4.0$. Fig.\
\ref{fig4}(b) shows that the AF channel correlations become
stronger with increasing coupling and the quasi-long-range
behavior develops for $J \gtrsim 2.0$.

In order to understand our numerical results for strong coupling,
we use the methods of Ref.\ \onlinecite{Sigrist92} to
derive an effective Hamiltonian valid for $t/J \ll 1$.
At $n_c=1$, the channel flavor
is the only degree of freedom and the
low-energy spectrum can be described by a pseudospin--$1/2$ model.
Accordingly, the
effective Hamiltonian is a Heisenberg model for the channel spin,
\begin{equation}
H = \frac{16 t^2}{3J}  \sum_i
\text{\boldmath{$\tau$}}_i \cdot \text{\boldmath{$\tau$}}_{i+1} \; ,
\label{eq:channelAF}
\end{equation}
where $\text{\boldmath{$\tau$}} = \frac{1}{2}\sum_{m m'} b^\dag_{i
m} \text{\boldmath{$\sigma$}}_{m m'}b^{\phantom{\dag}}_{i m'}$ is
defined in terms of hard-core bosons
\begin{equation}
\label{boson} b^{\dagger}_{i m}= \frac{1}{\sqrt{2}}
(c^{\dagger}_{i m \uparrow} f^{\dagger}_{i \downarrow} -
c^{\dagger}_{i m \downarrow} f^{\dagger}_{i \uparrow}) \; ,
\end{equation}
which represent local Kondo singlets. In one spatial dimension,
model (\ref{eq:channelAF}) has been solved exactly using the Bethe
ansatz.\cite{Bethe} The ground state is characterized by critical
AF correlations that decay as $1/r$, in agreement with our
numerical findings at quarter filling [see Fig.\ \ref{fig4}(a)].
This critical behavior is replaced by long-range AF ordering for
$D>1$. Due to the SU(2) pseudo-spin invariance of $H$, the AF
ordering is present for all three
$\text{\boldmath{$\tau$}}$-components. In particular, the order
along the $z$ axis corresponds to staggered orbital ordering while
the $x,y$-ordering gives rise to a Bose-Einstein condensation of
excitons (particle-hole singlet bound states between the two
channels).

For $n_c < 1$, the low-energy subspace in the strong-coupling
limit contains states with zero and one conduction
electron per site.
While the local state with one conduction electron can be
described by the hard-core bosons of Eq.\ (\ref{boson}),
the empty state acquires spin--$1/2$ character from the $f$
spins.
It can therefore be represented as a constrained fermion
$\varphi^{\dagger}_{i\sigma}= (1-n^{b}_i) f^{\dagger}_{i \sigma}$,
with $n^b_{i}=\sum_{m}b^{\dagger}_{i m} b^{\;}_{i m}$.
At strong coupling, the effective Hamiltonian has the
form
\begin{equation}
H_0 = -\frac{t}{2} \sum _{i \sigma,m} \left(
b^{\dag}_{i m} \varphi^{\phantom{\dag}}_{i \sigma}
\varphi^{\dag}_{i+1,\sigma} b^{\phantom{\dag}}_{i+1,m}
 + \mbox{H.c.} \right)
\; .
\nonumber
\end{equation}
In analogy to the infinite-$U$ Hubbard model, the ground-state wavefunction
can be written as a direct product of a charge, spin, and orbital
component
\begin{eqnarray}
|\psi_n\{\tau;\sigma\}\rangle&=& |n_c \rangle \otimes
|\tau_1 \cdots\; \tau_{L-N} \rangle \otimes |\sigma_1 \cdots\; \sigma_N
\rangle
\nonumber \\
=\sum_{i_1<i_2<...<i_N} && \!\!\!\!\!\!\!\!\!\!\!\!\!\!\!\!\!
\Gamma^{(n)}_{i_1 i_2 \cdots \; i_N} \varphi^{\dagger}_{i_1
\sigma_1} \cdots \; \varphi^{\dagger}_{i_N \sigma_N}
b^{\dagger}_{j_1\tau_1} \cdots \; b^{\dagger}_{i_{L-N}\tau_N} |0\rangle
\nonumber
\end{eqnarray}
where $\{\tau;\sigma\}=(\tau_1,...,\tau_{L-N};
\sigma_1,...,\sigma_{N})$ and $|0\rangle$ denotes the vacuum of
$\varphi$ and $b$ particles. The complete spin degeneracy of $H_0$ in
the strong coupling limit is lifted in ${\cal O} (t^2/J)$.
The effective Hamiltonian in this order contains only one term
that lifts this degeneracy:
\begin{equation}
H_1 = \frac{t^2}{2J} \sum _{i \sigma,m} \left(
\varphi^{\dagger}_{i+1,\sigma}
b^{\phantom{\dag}}_{i+1,m}
n^\varphi_i
b^{\dagger}_{i-1,m} \varphi^{\phantom{\dag}}_{i-1,\sigma}
+ \mbox{H.c.} \right) \; ,
\nonumber
\end{equation}
where $n^{\varphi}_i=\sum_{\sigma}\varphi^{\dagger}_{i \sigma}
\varphi^{\phantom{\dag}}_{i \sigma}$.
Here $H_1$ exchanges two spins by hopping of a fermion
over another to an empty next-nearest-neighbor site.
In the same way as for the
single-channel KLM,
it can be shown that the
off-diagonal elements of $H_1$ are all non-positive and the
Hamiltonian matrix in real-space is completely connected. The
Perron-Frobenius theorem (see, for instance,
Ref.~\onlinecite{matrix}) then states that the ground state is
unique and that the coefficients of the
wave-function can be chosen to be strictly positive.
The only spin state that has strictly positive coefficients in
each of the subspaces of $H_1$ is the one with maximum total spin.
Therefore, the ground state of the two-channel KLM below
quarter-filling is ferromagnetic in the strong-coupling limit,
in accordance with our numerical findings.

In this paper, we have mapped out the zero-tem\-per\-a\-ture phase
diagram of the two-channel KLM as a function of conduction band
filling $n_c$ and Kondo coupling strength $J$. Our main results
are that the phase diagrams of the two-channel and single-channel
KLMs are qualitatively similar for low band fillings ($n_c<1$ in
both models) but quite different at $n_c=1$. For $n_c<1$, the
addition of a second degenerate band of conduction electrons does
not alter the physical picture of the single-channel model if
$n_c$ is low or $J$ is large. The similarity between the two phase
diagrams for $n_c<1$ further suggests that the metallic ground
state of the two-channel KLM at weak coupling is determined by two
{\em independent\/} Kondo effects in both channels. In contrast,
at quarter-filling ($n_c=1$) the two-channel KLM exhibits a
quantum phase transition between this metallic phase and an
insulator characterized by strong AF correlations of the channel
degrees of freedom. Our perturbative analysis shows that this
insulating phase is present for any spatial dimension $D$ and has
long-range ordering for $D>1$. Therefore, we also expect a QCP
separating the metallic and the insulating phase for $D>1$. The AF
channel (or excitonic) fluctuations diverge at the QCP and can
produce a deviation from the normal Fermi-liquid behavior of the
metallic state. In this way we see that the deviations from the
Fermi-liquid behavior which are obtained in the dilute (impurity
model) and the concentrated (lattice model) limits have a common
origin in the fluctuations of the channel degree of freedom.

T.S., D.L.C., and C.D.B.\ acknowledge support from U.S.\ Department of
Energy.


\begin{thebibliography}{18}
\bibitem{FT-remark} The small negative contribution to $S_f(q)$ is due to
    the open boundary conditions in the DMRG simulations.
\end{thebibliography}
\end{document}